\documentclass[%
 reprint,
 superscriptaddress,
 amsmath,amssymb,
 aps,
]{revtex4-1}

\usepackage{graphicx}
\usepackage{dcolumn}
\usepackage{bm}
\usepackage{xcolor}
\usepackage{url}

\usepackage{subfigure}
\usepackage{multirow}

\begin{document}


\title{Simulation of the neutron-tagged deep inelastic scattering at EicC}

\author{Gang Xie}
\affiliation{Guangdong Provincial Key Laboratory of Nuclear Science, Institute of Quantum Matter, South China Normal University, Guangzhou 510006, China}
\affiliation{Institute of Modern Physics, Chinese Academy of Sciences, Lanzhou 730000, China}

\author{Mengyang Li}
\affiliation{Guangdong Provincial Key Laboratory of Nuclear Science, Institute of Quantum Matter, South China Normal University, Guangzhou 510006, China}
\affiliation{Institute of Modern Physics, Chinese Academy of Sciences, Lanzhou 730000, China}

\author{Chengdong Han}
\affiliation{Institute of Modern Physics, Chinese Academy of Sciences, Lanzhou 730000, China}
\affiliation{School of Nuclear Science and Technology, University of Chinese Academy of Sciences, Beijing 100049, China}

\author{Rong Wang}
\email{rwang@impcas.ac.cn}
\affiliation{Institute of Modern Physics, Chinese Academy of Sciences, Lanzhou 730000, China}
\affiliation{School of Nuclear Science and Technology, University of Chinese Academy of Sciences, Beijing 100049, China}

\author{Xurong Chen}
\email{xchen@impcas.ac.cn}
\affiliation{Guangdong Provincial Key Laboratory of Nuclear Science, Institute of Quantum Matter, South China Normal University, Guangzhou 510006, China}
\affiliation{Institute of Modern Physics, Chinese Academy of Sciences, Lanzhou 730000, China}
\affiliation{School of Nuclear Science and Technology, University of Chinese Academy of Sciences, Beijing 100049, China}


\date{\today}

\begin{abstract}
Measuring the pionic structure function is of high interests
as it provides a new area for understanding the strong interaction among quarks
and testing the QCD predictions.
To this purpose, we investigate the feasibility and the expected impacts
of a possible experiment on EicC.
We show the simulation results on the statistical precision of an EicC measurement,
based on the model of leading neutron tagged DIS process
and the parton distribution functions of the pion from JAM18 global analysis.
The simulation shows that at EicC, the kinematics cover $x_{\pi}$ range from 0.01 to 1,
and $Q^2$ range from 1 GeV$^2$ to 50 GeV$^2$, within the acceptable statistical uncertainty.
Assuming an integrated luminosity of 50 fb$^{-1}$, in the low-$Q^{2}$ region ($<10$ GeV$^2$),
the MC data show that the suggested measurement in the whole $x_{\rm\pi}$ range reaches very high precision ($<3$\%).
To perform such an experiment, only the addition of a far-forward neutron calorimeter is needed.
\end{abstract}

\maketitle

\section{Introduction}
\label{sec:intro}

Pion, the lightest hadron made of the first-generation quark and antiquark,
plays a fundamental role in particle and nuclear physics, as the long-range
nuclear force carrier which binds the nucleons together into a nucleus \cite{Yukawa:1935xg}.
In theory, it is a good approximation of the Nambu-Goldstone boson \cite{Nambu:1960tm,Goldstone:1962es}
from the spontaneous chiral symmetry breaking,
however the emergence of its small mass (much smaller than that of the proton)
is not yet understood quantitatively and experimentally \cite{Roberts:2019ngp,Roberts:2020udq,Cui:2020dlm,Chen:2020ijn}.
Recent progresses from Dyson-Schwinger equations (DSE), which is a nonperturbative
quantum chromodynamics (QCD) approach, show that the dressed quark mass which
comes from the dynamical chiral symmetry breaking \cite{Roberts:1994dr,Hawes:1993ef,Maris:1997tm} is greatly cancelled by
the attraction interaction between the quark and the antiquark \cite{Roberts:2019ngp,Roberts:2020udq,Maris:1997hd}.
Understanding the properties of the simplest hadron from its structure
is a remarkable advancement in revealing the strong interaction.

Along with the emergence of the pion mass, DSE predicts a broadening
parton distribution amplitude (PDA) \cite{Chang:2013nia,Shi:2015esa,Chen:2018rwz}, compared to the asymptotic form of PDA
by perturbative QCD \cite{Lepage:1980fj,Farrar:1979aw,Efremov:1979qk}. The width of pion quark distribution
also becomes wider at the hadronic scale $Q_0^2$ (a very low scale where
sea quarks and gluons disappear). Using a renormalization-group-invariant
process-independent strong coupling, the initial quark distributions at $Q_0^2$
is connected to the extracted parton distribution functions (PDF) at high $Q^2$ in experiment
\cite{Hecht:2000xa,Holt:2010vj,Aicher:2010cb,Nguyen:2011jy,Chang:2014lva,Chang:2014gga,Ding:2019lwe}.
Moreover the predicted valence quark distribution from dynamical chiral symmetry breaking
has the similar large-$x$ ($x\rightarrow 1$) behavior of the perturbative QCD predictions \cite{Ezawa:1974wm,Farrar:1975yb,Berger:1979du}.
Measuring the pion structure in the full range of $x$ and a broad range of $Q^2$
provides a promising window to see the dynamical chiral symmetry breaking,
which is one of the prominent features of QCD theory, and to uncover the related emergence
phenomena of the strong interaction.

Experimentally, the collinear parton distributions of nucleon
are measured very precisely with the helps of the high energy accelerator facilities worldwide,
however we have far less experimental data about the parton distributions of pion.
Measuring the pionic structure is not easy, since there is no pion target in experiment as it decays quickly.
All the data on pion valence quark distributions are measured in the Drell-Yan reaction
induced by the pion beam \cite{Badier:1980jq,Betev:1985pf,Bordalo:1987cs,Conway:1989fs},
more than thirty years ago.
Recently, by exploiting the ``pion cloud'' around the proton, the pion structure function
at small $x_{\rm \pi}$ ($\lesssim 0.01$) is analyzed from the leading neutron (LN) tagged deep inelastic scattering (DIS) data
at HERA collider \cite{Chekanov:2002pf,Aaron:2010ab}. Therefore, to fill the data gap in the range $0.01\lesssim x_{\rm \pi} \lesssim 0.2$
is of particular interest on the experimental side.
More data in the full $x$ range and at different $Q^2$ are still scarce.
Furthermore, measuring the pion valence quark distributions at large $x$ using the LN-DIS technique,
and comparing it with that from the Drell-Yan process will reinforce our understanding
of the perturbative QCD theory on the dynamics when $x_{\rm\pi}$ approaches one.
Lastly, the experimental data from sea quark region to valence quark region will definitely
provide an opportunity to test various theoretical approaches,
such as DSE \cite{Ding:2019lwe,Cui:2020dlm}, lattice QCD \cite{Detmold:2003tm,Shugert:2020tgq,Sufian:2020vzb,Gao:2020ito},
holographic QCD \cite{deTeramond:2018ecg}, light-front quantization \cite{Lan:2019vui},
chiral quark model \cite{Suzuki:1997wv,Nam:2012vm,Nematollahi:2018iac}, constituent quark model \cite{Szczepaniak:1993uq,Frederico:1994dx,Watanabe:2016lto},
QCD sum rule \cite{Ioffe:1999hz}, and the dynamical parton model prediction
with a naive nonperturbative input \cite{Lou:2015nta,Han:2018wsw}.

Recently the JAM Collaboration \cite{Barry:2018ort}
and xFitter Collaboration \cite{Novikov:2020snp} performed the global analyses
of the pionic parton distribution functions from the Drell-Yan data and the LN-DIS data.
JAM's analysis shows that the addition of the LN-DIS data constrains better the gluon distribution.
However, compared to the fit with only the Drell-Yan data,
the valence quark, sea quark and gluon distributions change sizely  \cite{Barry:2018ort}.
Therefore more measurements on the LN-DIS process are necessary for the future global fit studies.
The xFitter Collaboration finds that the current experimental data are not enough
to unambiguously determine the sea quark and gluon distributions of the pion.
Moreover xFitter Collaboration studied the model uncertainties that are related to the variations
of the factorization and renormalization scales and the flexibility of the chosen parametrization.
They estimate that these model uncertainties are significant  \cite{Novikov:2020snp}.
Hence more experimental data on the pion structure in the future are very critical
to constrain the sea quark and gluon distributions of the pion and
to understand the model uncertainties in the global fit.

Now, there are some heating discussions on building a low energy electron-ion collider
in China (EicC), by upgrading the under-construction high-intensity heavy ion accelerator
facility (HIAF) \cite{Chen:2018wyz,EicCWhitePaperChinese}.
By using the same method verified at HERA, EicC with the center-of-mass (c.m.)
energy about 20 GeV provides a competing opportunity to acquire the pionic structure
data in the range $0.01\lesssim x_{\rm\pi} \lesssim 1$. Hence in this work, we investigate
the feasibility and the kinematical distributions of interests,
providing some guidances for detecting the final-state particles.
Then we focus on the anticipated statistical errors of pionic structure function
for a suggested LN-DIS experiment at EicC.

The organization of the paper is as follows. The formulae of the
LN-DIS process to study the structure of pion are discussed in Sec. {\ref{sec:LNDIS}}.
The PDFs of the pion from JAM18 global analysis used in this simulation
and the dynamical PDF model of the pion
are introduced in Sec. {\ref{sec:pionPDFs}}.
The commonly used invariant kinematical distributions of the reaction
and the angular distributions of the final-state particles
are given in Sec. {\ref{sec:KineDistributions}}, for the proposed experiment at EicC.
Then the statistical error projections of the pionic structure function $F_2$
are shown for an assumed experimental run of 50 fb$^{-1}$ integrated luminosity in Sec. {\ref{sec:F2piErrors}}.
At the end, we give some discussions and a concise summary in Sec. {\ref{sec:summary}}.

\section{Leading-neutron DIS and pionic structure function}
\label{sec:LNDIS}

\begin{figure}[htbp]
\centering
\includegraphics[scale=0.42]{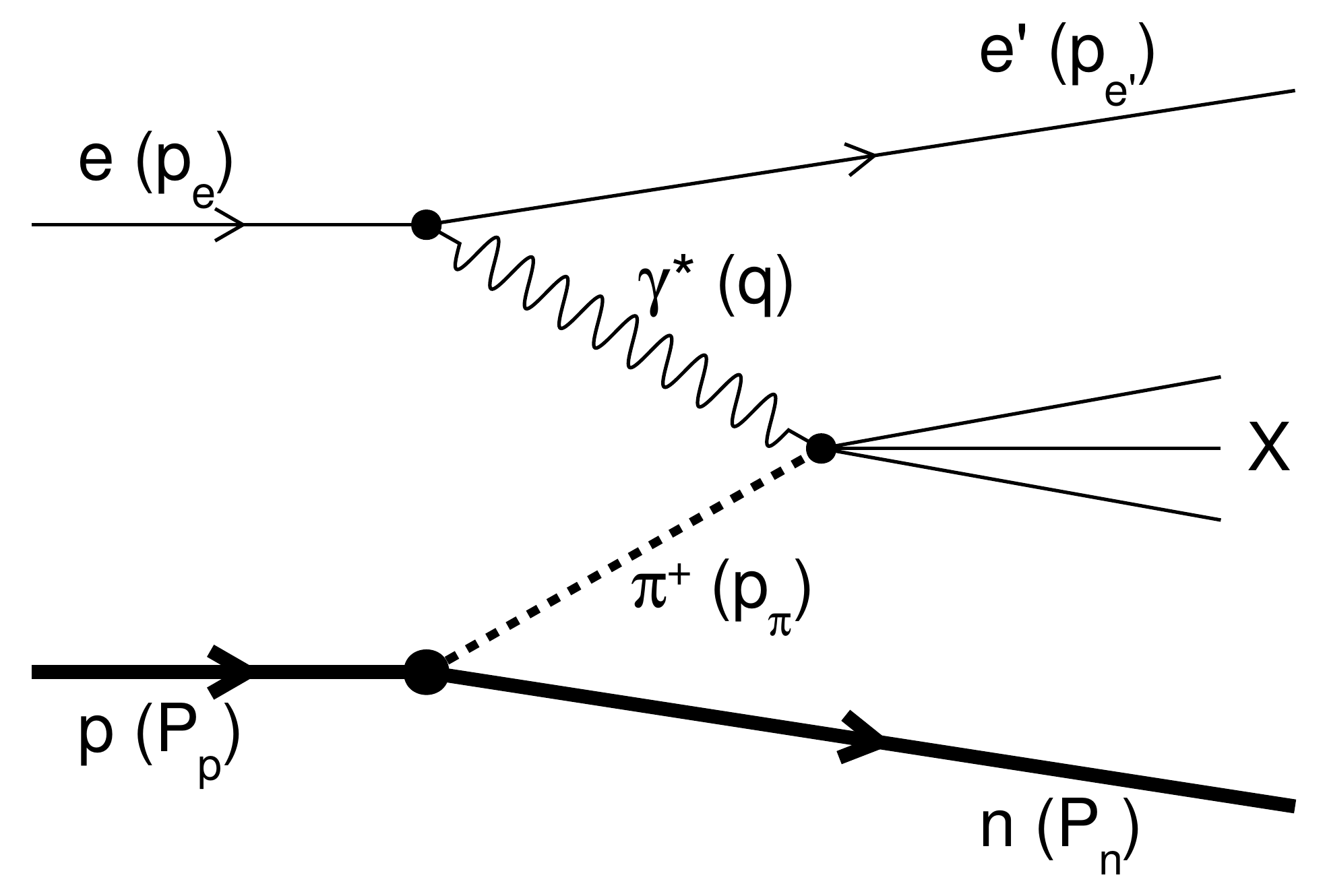}
\caption{The Sullivan process \cite{Sullivan:1971kd} for the leading-neutron deep inelastic scattering,
where the one-pion exchange process dominates. }
\label{fig:LNDIS-diagram}
\end{figure}

To explore the structure of the pion in e-p scattering,
the key idea is to take the advantage of the abundant ``pion cloud'' around the proton.
The $\pi^+$ in the n-$\rm\pi^+$ Fock state of the proton dissociation \cite{Holtmann:1996ac}
is abundant because of the large $\pi-N-N$ coupling.
Fig. \ref{fig:LNDIS-diagram} shows a schematic diagram of the LN-DIS,
where the exchanged pion of Sullivan process \cite{Sullivan:1971kd}
is probed and broke up by the virtual photon.
In the case of e-$\rm\pi^+$ DIS, the neutron spectator
carries a large fraction of the momentum of the incoming proton and a small transverse momentum $P_{T,n}$.
The final neutron in this case has a large longitudinal momentum and rapidity, which is called
the leading neutron and quite distinguishable from the neutron fragment from the normal DIS \cite{Chekanov:2002pf,Aaron:2010ab}.
More theoretical calculations indicate that the structure of the virtual pion at low virtuality ($|t=m_{\pi}^2|<0.6$ GeV$^2$)
can be effectively extrapolated to the on-shell pion, based on an analysis of the Bethe-Salpeter amplitude \cite{Qin:2017lcd}.

According to the momenta labeled in Fig. \ref{fig:LNDIS-diagram}, the commonly used
virtuality of the photon probe $Q^2$, the Bjorken variable $x_B$,
and the inelasticity $y$ of DIS process are defined as,
\begin{equation}
Q^2 \equiv -q^2, ~ x_{\rm B} \equiv \frac{Q^{2}}{2P_{\rm p} \cdot q}, ~ y \equiv \frac{P_{\rm p} \cdot q}{P_{\rm p} \cdot P_{\rm e}}.
\label{eq:DIS-kine-definition}
\end{equation}
Another kinematical variables related to the final-state neutron
are the longitudinal momentum fraction $x_L$ and the square of the momentum transfer to the virtual pion, $t$,
\begin{equation}
x_{\rm L} \equiv \frac{P_{\rm n}\cdot q}{P_{\rm p}\cdot q}, ~ t \equiv (P_{\rm p} - P_{\rm n})^2 = p_{\rm \pi^{*}}^2.
\label{eq:neutron-kines}
\end{equation}
$x_{\rm L}$ is the longitudinal momentum fraction (energy fraction approximately)
of the final neutron to the incoming proton.
In experiment, the LN-DIS process dominates in the large-$x_L$ region ($\gtrsim 0.5$) \cite{Aaron:2010ab},
hence the proper cut on the $x_L$ variable is the efficient way
to select the events that is sensitive to the pion structure.
Viewing the virtual pion as the effective target, similar to the definition of normal Bjorken variable,
the momentum fraction of the parton inside the pion is given by,
\begin{equation}
x_{\rm \pi} \equiv \frac{Q^2}{2p_{\rm \pi}\cdot q} = \frac{x_{\rm B}}{1-x_{\rm L}}.
\label{eq:neutron-kines}
\end{equation}
From the above definition, we see that the smallest momentum fraction of the parton in the pion
measured in LN-DIS is larger than the smallest momentum fraction of the parton in the proton
measured in DIS, for the e-p collisions with the same c.m. energy.

To estimate the statistics of LN-DIS events and the distributions of the kinematical observables at EicC,
we need to calculate the differential cross-section of the channel.
Integrating the azimuthal angles, the four-fold differential cross-section of LN-DIS process
is expressed with the semi-inclusive structure function $F_2^{\rm LN(4)}(Q^2, x_{\rm B}, x_{\rm L}, t)$ \cite{Holtmann:1996ac,Chekanov:2002pf,Aaron:2010ab},
\begin{equation}
\begin{split}
&\frac{d^4\sigma({\rm ep\rightarrow enX})}{dx_{\rm B}dQ^2dx_{\rm L}dt} =
\frac{4\pi\alpha^2}{x_{\rm B}Q^4}\left(1-y+\frac{y^2}{2}\right)F_2^{\rm LN(4)}(Q^2, x_{\rm B}, x_{\rm L}, t)\\
&= \frac{4\pi\alpha^2}{x_{\rm B}Q^4}\left(1-y+\frac{y^2}{2}\right) F_2^{\rm \pi}\left(\frac{x_{\rm B}}{1-x_{\rm L}} ,Q^2\right)
f_{\rm \pi^+/p}(x_{\rm L},t).
\end{split}
\label{eq:DiffXSection-four}
\end{equation}
In the above formula, the leading-neutron structure function $F_2^{\rm LN(4)}$
is then factorized into the pionic structure function $F_2^{\rm \pi}$
and the pion flux around the proton $f_{\rm \pi^+/p}$.
The pion flux is usually evaluated to be a pion pole in the effective field theory \cite{Holtmann:1996ac,Chekanov:2002pf,Aaron:2010ab},
\begin{equation}
\begin{split}
& f_{\rm \pi^+/p}(x_{\rm L},t)=\\
& \frac{1}{2\pi}\frac{g^2_{\rm pn\pi}}{4\pi}(1-x_{\rm L})\frac{-t}{(m_{\pi}^2-t)^2}{\rm exp}\left(R^2_{\rm n\pi}\frac{t-m_{\pi}^2}{1-x_{\rm L}}\right),
\end{split}
\label{eq:pion-flux}
\end{equation}
where $g^2_{\rm pn\pi}/4\pi=13.6$ is the ${\rm \pi}-N-N$ effective coupling,
and $R_{\rm n\pi}=0.93$ GeV$^{-1}$ is an adjustable parameter describing the radius of n-${\rm \pi}$ Fock state \cite{Holtmann:1996ac}.
By integrating over the $t$ variable, the three-fold LN structure function is
also used often, which is written as,
\begin{equation}
\begin{split}
F_2^{\rm LN(3)}(Q^2, x_{\rm B}, x_{\rm L}) = \int_{t_1}^{t_0} F_2^{\rm LN(4)}(Q^2, x_{\rm B}, x_{\rm L}, t) dt.
\end{split}
\label{eq:F2-LN3}
\end{equation}

For now, the theoretical framework for the pion structure function measurement in e-p process
is mature and has been tested with the pioneering experiments on the HERA facility.
The shape of the structure function of the pion is encoded in the LN structure function.
So for completing the quantitative calculation of the cross-section,
the last thing left is to seek for a valid structure function model of the pion
in a wide kinematical range of $x_{\pi}$ and $Q^2$.

\section{Parton distribution functions of the pion}
\label{sec:pionPDFs}

In this simulation, we take the pion PDF of the recent JAM18 global analysis
as the input for the cross section model.
To check JAM18 PDF of the pion and to see the status of our understanding
on pionic structure, the JAM18 global fit result is compared
with the experimental data and the dynamical parton model \cite{Han:2018wsw,piIMParton-github}.
The pionic PDF based on the dynamical parton model is called as piIMParton
for the convenience of discussions and the reference \cite{Han:2018wsw,piIMParton-github}.
The salient argument of the dynamical parton model is
that the nonperturbative input consists of only the valence quark distributions
at extremely low $Q_0^2$ scale.
The low $Q_0^2$ scale is estimated to be around 0.1 GeV$^2$,
which is also called the hadronic scale, since at this scale
only the minimum components (valence) of the hadron can be resolved.
In this dynamical parton model, all the sea quarks and gluons are produced
from the parton splitting processes governed by the DGLAP equations \cite{Altarelli:1977zs}.
The dynamical parton model is also called the radiative parton model,
as all the sea quarks and gluons are given by the QCD radiations.
Hence it is very interesting to compare the JAM18 PDF with the piIMParton PDF.

\begin{figure}[htbp]
\centering
\includegraphics[scale=0.39]{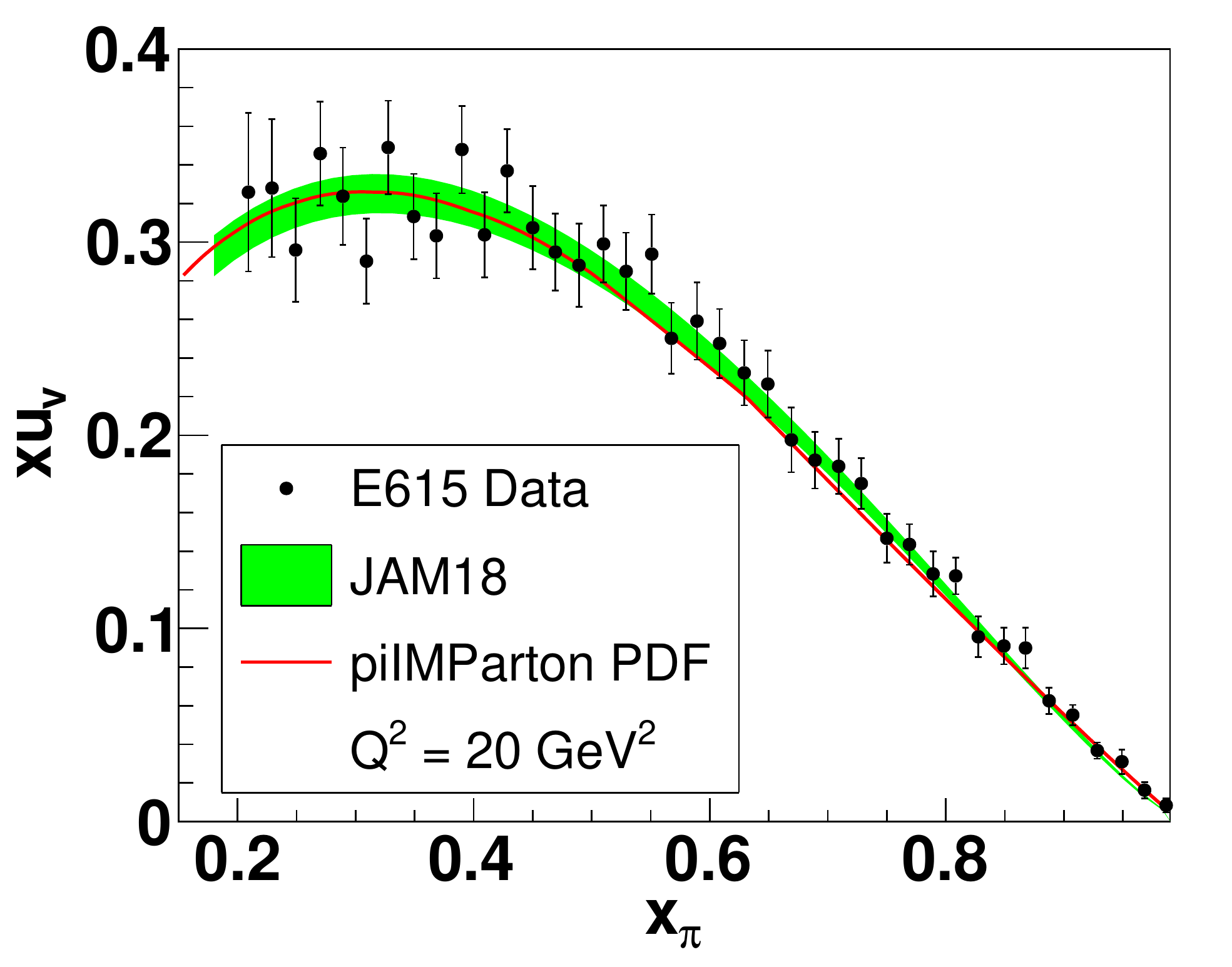}
\caption{The comparisons among JAM18 PDF of the pion, the dynamical valence quark distribution from piIMParton \cite{piIMParton-github},
and the E615 experimental data \cite{Conway:1989fs} from pion-induced Drell-Yan reaction.
The green band shows one $\sigma$ uncertainty of JAM18 global analysis. }
\label{fig:pion-valence-distribution}
\end{figure}

\begin{figure}[htbp]
\centering
\includegraphics[scale=0.45]{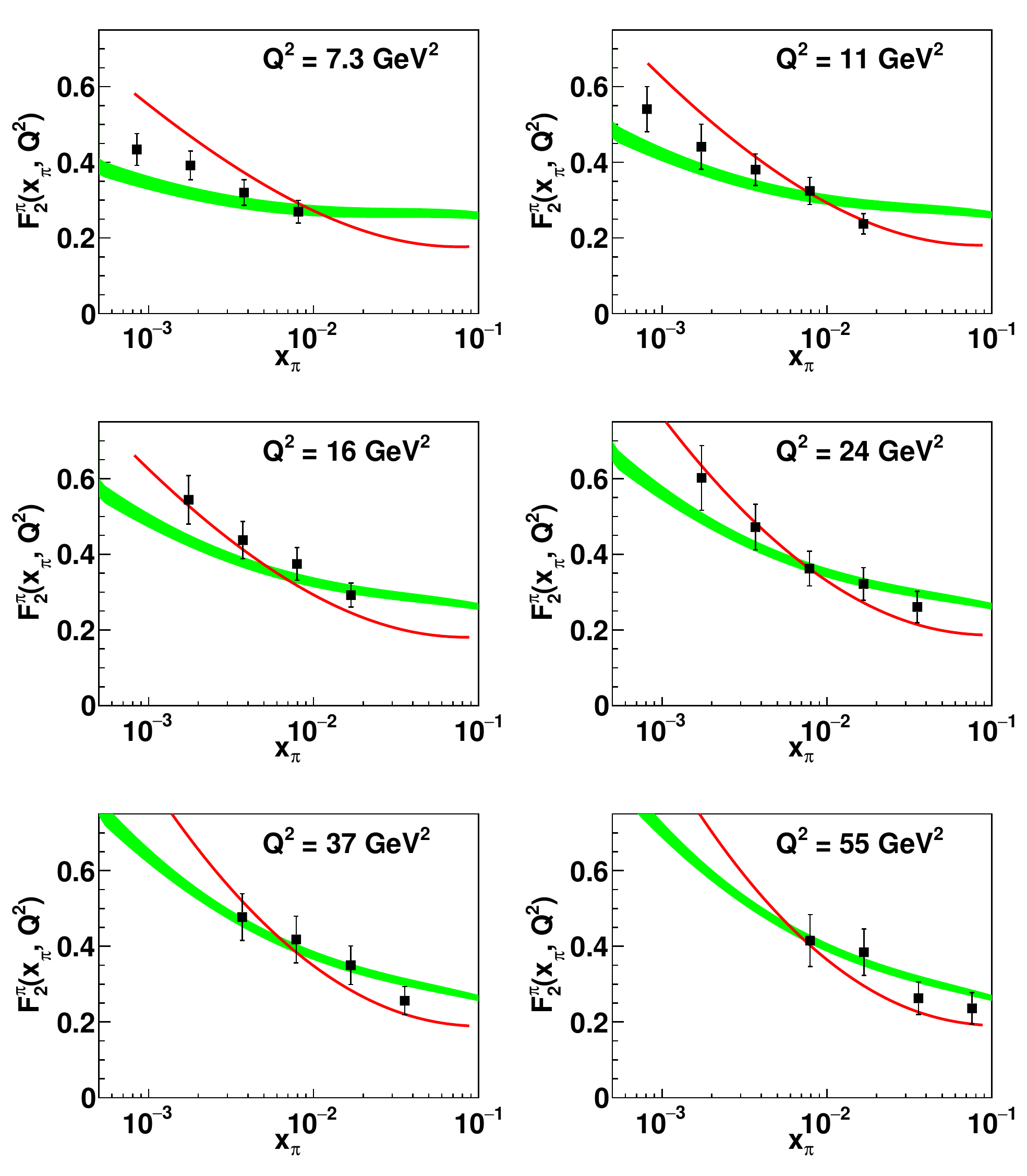}
\caption{The comparisons among the global fit by JAM18 (green bands),
the predictions from piIMParton PDF of the pion (red solid curves),
and the experimental data of the pionic structure function measured by H1 Collaboration (black squares).
The green bands show one $\sigma$ uncertainties of JAM18 global analysis.
The pionic structure function extracted by H1 is from an analysis
of the LN-DIS data in the kinematical region of $x_{\rm L}$ around 0.73 \cite{Aaron:2010ab}. }
\label{fig:F2pi-H1Data}
\end{figure}

Fig. \ref{fig:pion-valence-distribution} shows the valence quark distribution of the pion
from JAM18 global analysis and piIMParton \cite{Han:2018wsw,piIMParton-github},
compared with that extracted from the ${\rm \pi}$-nucleus Drell-Yan data
by E615 Collaboration \cite{Conway:1989fs}.
We can see the excellent agreements among the JAM18 global fit, the dynamical parton model calculation (piIMParton),
and the experimental data in the range $0.2<x_{\rm \pi}<1$.
We also notice that in the region $x_{\rm\pi} \lesssim 0.6$
the experimental data exhibit big uncertainties.
So the aim of future experiments is to collect more data and
to reduce much the uncertainties in the small-$x_{\rm \pi}$ region.
In the small-$x_{\rm \pi}$ region ($x_{\rm \pi}\lesssim 0.01$),
there are a few experimental data of pion structure function obtained
in the H1 and ZEUS experiments at HERA \cite{Chekanov:2002pf,Aaron:2010ab}.
Fig. \ref{fig:F2pi-H1Data} shows the comparisons among
the structure function prediction from JAM18 PDF,
the piIMParton model prediction, and the structure function data by H1.
Note that in the model calculation of the structure function
$F_2^{\rm \pi}$, only u, d, and s quark contributions are taken into account.
Amazingly, the calculations at small $x_{\rm \pi}$ of both the JAM18 pionic PDF
and the dynamical sea quark distributions
are consistent with the up-to-date experimental measurements of pion structure.
Some differences are found between the JAM18 prediction and the piIMParton prediction.
Therefore the future experimental measurements are important to well distinguish the models.

\begin{figure}[htbp]
\centering
\includegraphics[scale=0.45]{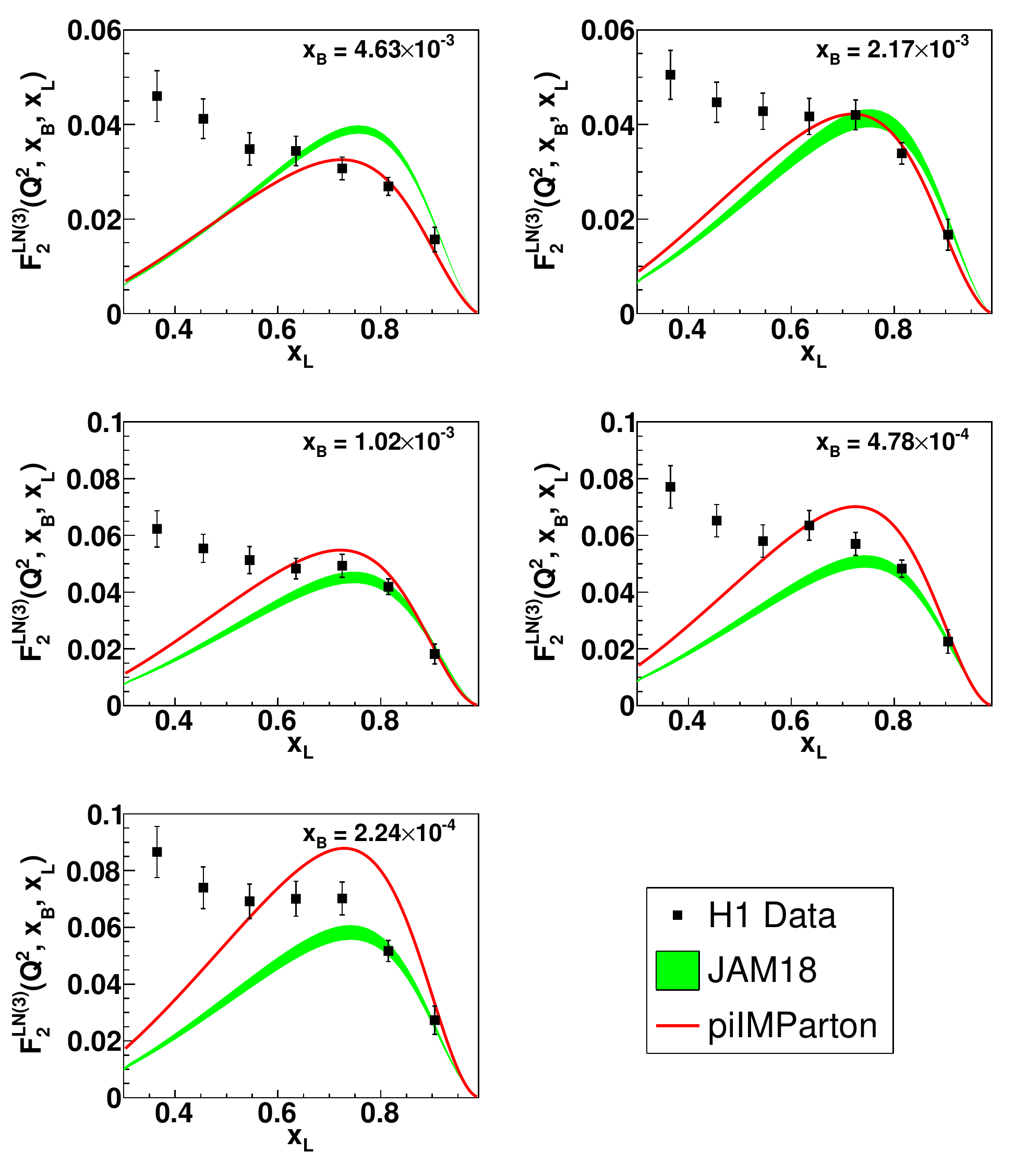}
\caption{The comparisons among the predictions of the LN structure function from JAM18 PDF,
the predictions from piIMParton model \cite{piIMParton-github},
and the H1 data \cite{Aaron:2010ab}, at $Q^2=11$ GeV$^2$.
The green bands show one $\sigma$ uncertainties of JAM18 global analysis. }
\label{fig:F2LN3-H1Data}
\end{figure}

\begin{figure}[htbp]
\centering
\includegraphics[scale=0.39]{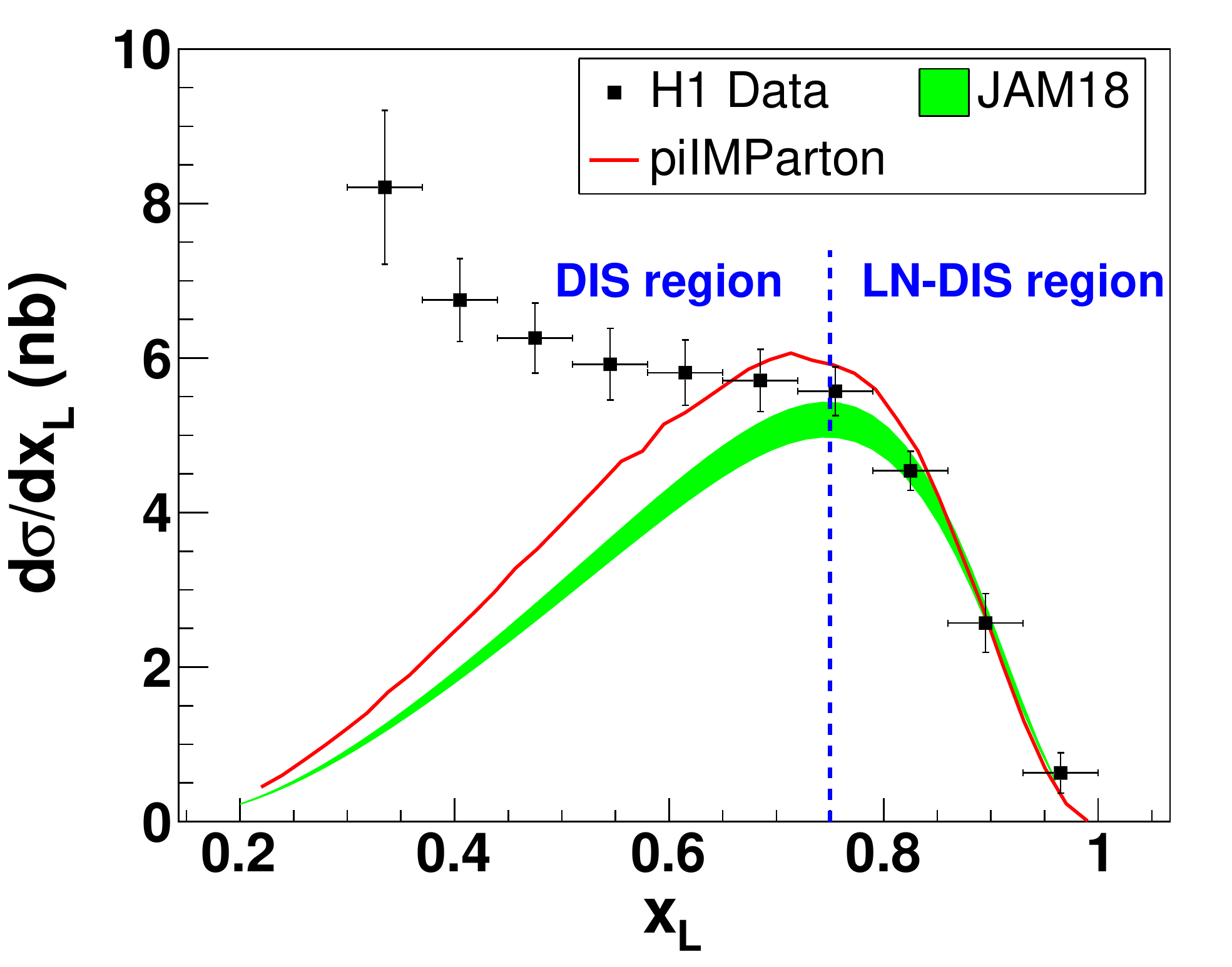}
\caption{The comparisons among the predictions of one-fold differential cross-section from JAM18 PDF,
the predictions from piIMParton model \cite{piIMParton-github}, and the H1 data \cite{Aaron:2010ab},
integrated in the kinematical range of $6<Q^2<100$ GeV$^2$,
$1.5\times 10^{-4} <x_{\rm B} <3\times 10^{-2}$, and $P_{\rm T}^{\rm n}<0.2$ GeV/c.
The green band shows one $\sigma$ uncertainty of JAM18 global analysis.
In the LN-DIS dominant region (eg. $x_{\rm\pi}>0.75$), the cross-section can be explained with
the e-${\rm \pi}$ DIS formula and the dynamical parton distribution functions of the pion. }
\label{fig:dsigma_dxL-H1Data}
\end{figure}

To check carefully the pionic PDF of JAM18 and
the dynamical parton distributions of piIMParton,
we calculate more experimental observables
and compare them to the recent H1 experiment \cite{Aaron:2010ab}.
Fig. \ref{fig:F2LN3-H1Data} shows the comparisons among the calculated three-fold
leading-neutron structure functions $F_2^{\rm LN(3)}$ and the H1 data.
In the large-$x_{\rm L}$ region ($x_{\rm L}\gtrsim 0.6$),
the model calculations are well consistent with the experimental data.
This is because the e-${\rm \pi^{*}}$ DIS process plays an important
role in the regime of large $x_{\rm L}$.
In the small-$x_{\rm L}$ region, if the contribution from the normal e-p DIS process
is added in the model calculation, then the experimental data is well explained and described.
Nevertheless, the focus of this simulation is on the region of $x_{\rm L}>0.75$.
We also calculate the differential cross-section as a function of $x_{\rm L}$,
which is shown in Fig. \ref{fig:dsigma_dxL-H1Data} compared to the H1 data.
In the leading-neutron production region, the predictions of both JAM18 PDF
and the dynamical parton model agree well with the H1 experiment at HERA.
The validity of our LN-DIS cross-section is verified with an independent calculation by
RAPGAP software. In the literature \cite{Aaron:2010ab}, it is shown that the H1 data in the small-$x_{\rm L}$ region
is well explained by the normal DIS process on the proton. The combination of the LN-DIS process
and the normal DIS process with neutron fragment reproduces the experimental data
in the whole range of $x_{\rm L}$ \cite{Aaron:2010ab}.

\begin{figure}[htbp]
\centering
\includegraphics[scale=0.42]{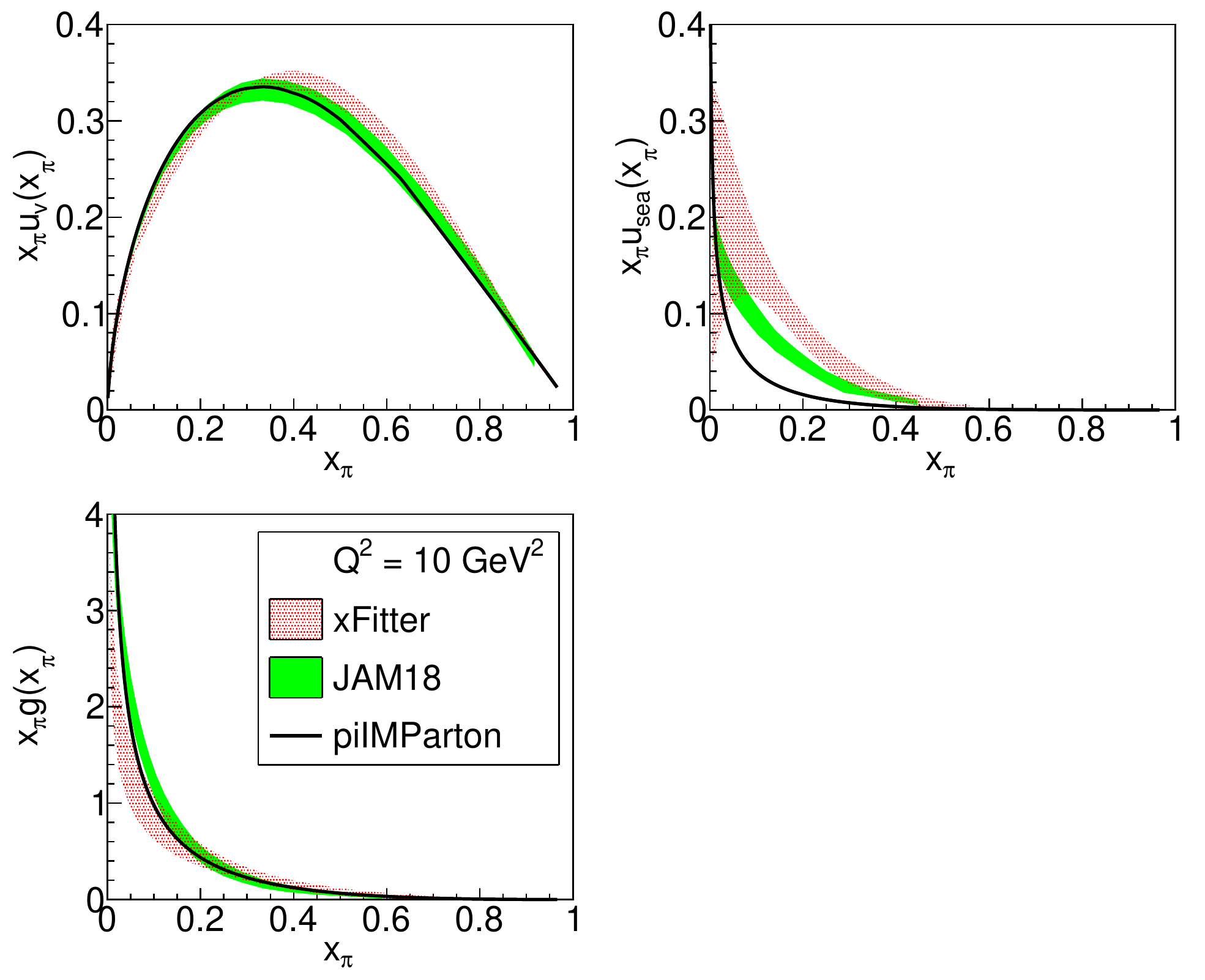}
\caption{The comparisons among the global analysis by JAM Collaboration \cite{Barry:2018ort},
the global analysis by xFitter Collaboration \cite{Novikov:2020snp},
and piIMParton PDFs from the dynamical parton model.
The green bands show one $\sigma$ uncertainties of JAM18 global analysis.
The red bands show one $\sigma$ uncertainties of xFitter global analysis.
The PDF data of the global analyses (JAM18 and xFitter) are taken from LHAPDF \cite{Buckley:2014ana}. }
\label{fig:Pion_PDF_Sets}
\end{figure}

Fig. \ref{fig:Pion_PDF_Sets} shows the comparisons between
the pionic parton distributions from the recent global analyses
of the current experimental data \cite{Barry:2018ort,Novikov:2020snp}
and the dynamical parton distributions of piIMParton.
We can see that there are some differences among the different PDF sets.
Nonetheless the parton distributions of the pion from JAM18 \cite{Barry:2018ort}
are consistent with the xFitter global fit \cite{Novikov:2020snp}
and the dynamical parton distributions of the pion \cite{Han:2018wsw,piIMParton-github}.
The purely dynamical sea quark distribution of piIMParton is lower than
that from the phenomenological studies of global analyses,
since in the dynamical parton model there is no intrinsic sea quark distributions at $Q_0^2$
of an arbitrary parametrization.
So far the pion PDFs from the global analyses exhibit big uncertainties
due to the inadequate experimental data at small $x_{\rm \pi}$.
Moreover there are the model uncertainties from varying the factorization or renormalization
scales and the flexibility of parametrization of the input parton distributions.
Therefore there is no significant difference of using
whatever model of the pionic parton distribution functions.

Right now there is no measurement of LN-DIS at EicC energy.
Therefore we use the JAM18 PDF of the pion and the one-pion
exchange model to give projections of a possible EicC experiment.
Every Monte-Carlo simulation has some model uncertainties.
From the above demonstrations, we see that the pion PDFs from JAM18 \cite{Barry:2018ort}
are acceptable to reproduce the
E615 data and H1 data at very high energies.
Convincingly, the cross-section model and the pionic parton distributions
used in this simulation bring the controlled uncertainties
for the projections of a suggested LN-DIS experiment at EicC.

\section{Distributions of the invariant and final-state kinematics}
\label{sec:KineDistributions}

Following the theoretical framework described in Sec. {\ref{sec:LNDIS}} and Sec. {\ref{sec:pionPDFs}},
we developed a Monte-Carlo (MC) simulation program,
which generates the LN-DIS events very efficiently.
In the simulation, the electron beam energy is 3.5 GeV
and the proton beam energy is 20 GeV,
which is at a typical collision energy for the future EicC \cite{Chen:2018wyz,EicCWhitePaperChinese}.
Inside the phase space of the LN-DIS process,
we apply the following kinematical ranges
for the MC sampling: $0.01$ GeV$^2$ $<-t<1$ GeV$^2$,
$0.5<x_{\rm L}<1$, $x_{\rm B,min}<x_{\rm B}<1$,
1 GeV$^2$ $<Q^2<$ 50 GeV$^2$, and $W^2>4$ GeV$^2$,
to focus on the kinematical region of interests.

\begin{figure}[htbp]
\centering
\includegraphics[scale=0.43]{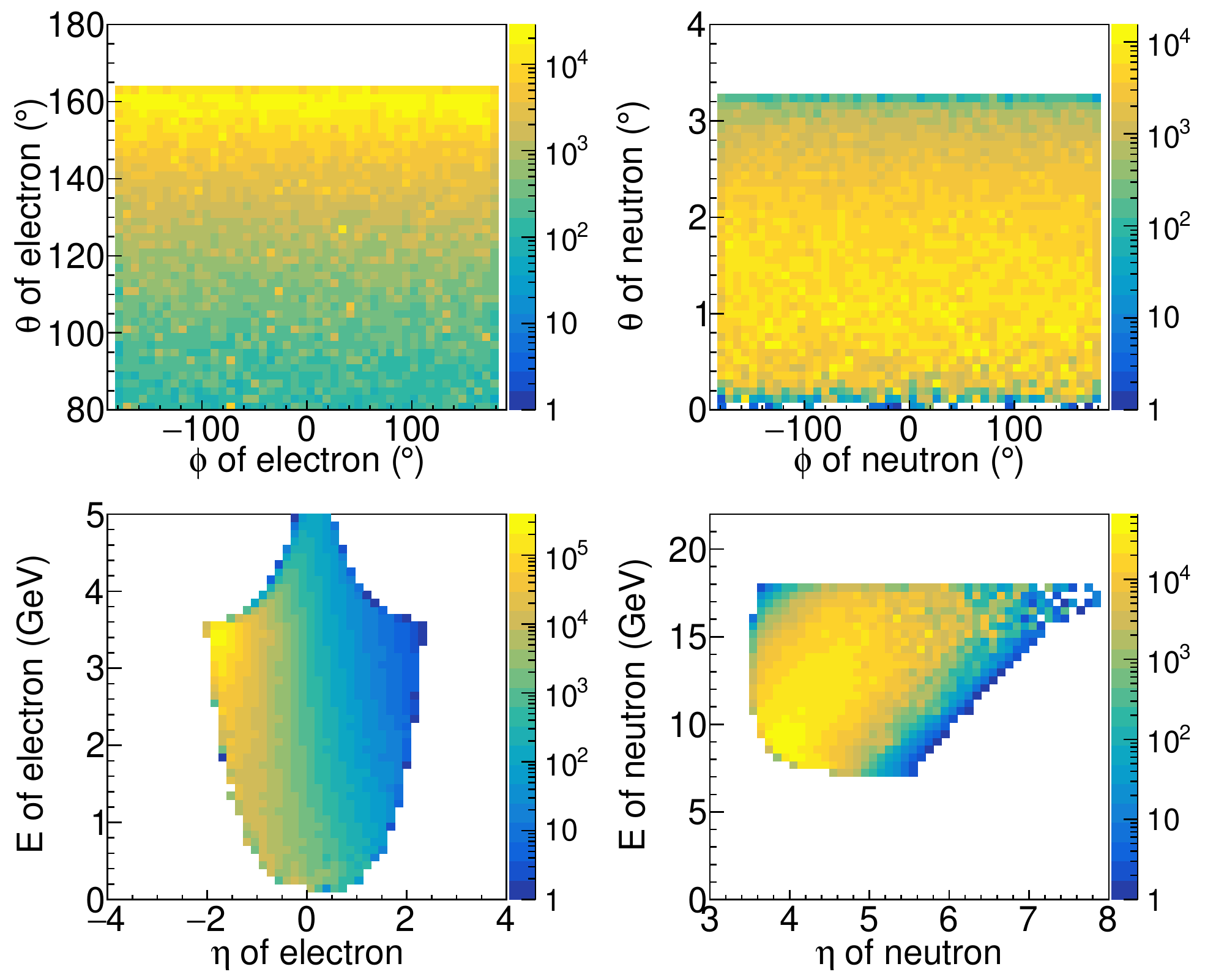}
\caption{The cross-section weighted kinematical distributions of the final-state particles
in the MC simulation. The angular, energy and pseudo-rapidity distributions are shown. }
\label{fig:final-state-kine}
\end{figure}

\begin{figure}[htbp]
\centering
\includegraphics[scale=0.43]{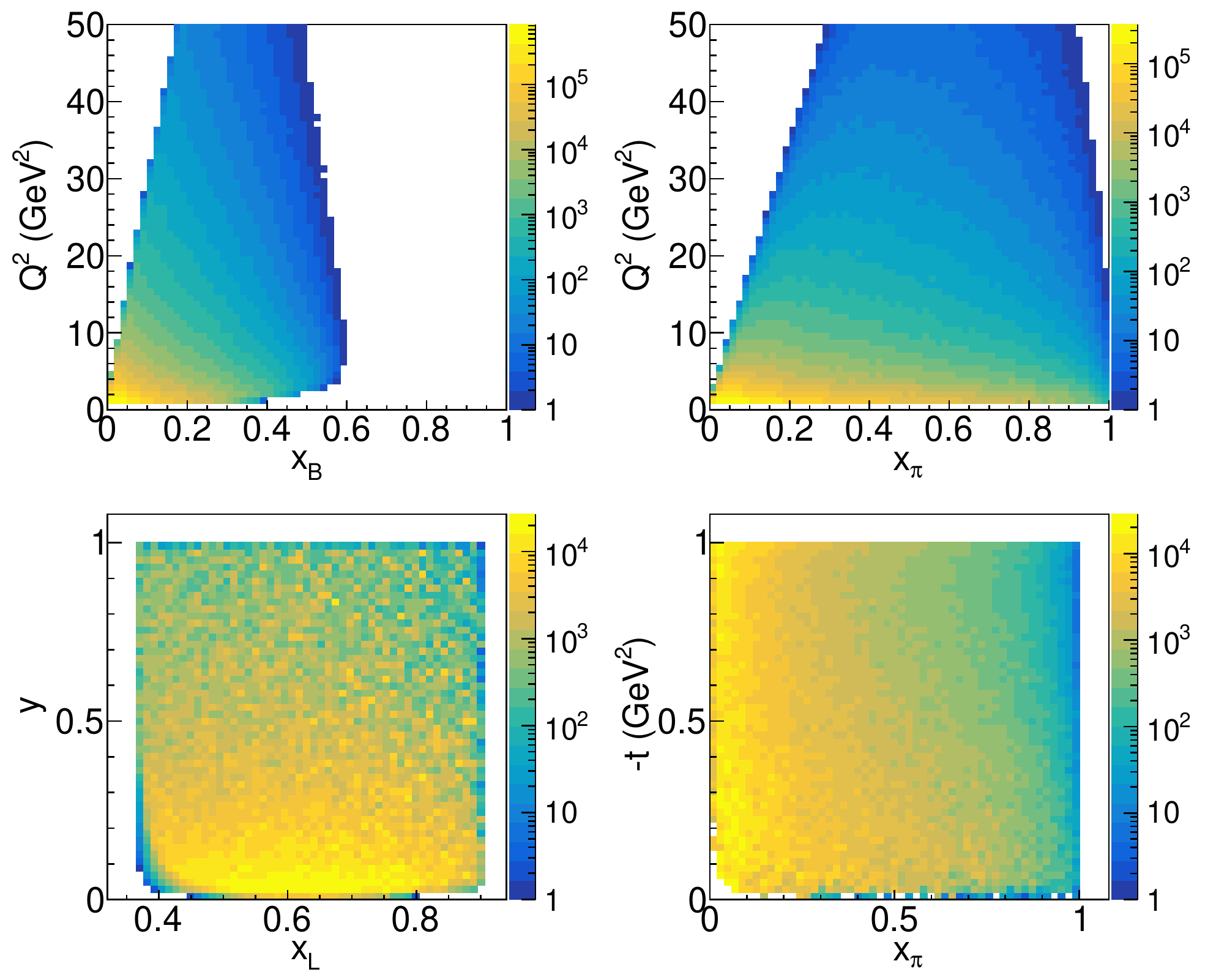}
\caption{The cross-section weighted distributions of the invariant kinematics
in the MC simulation. The definitions of the invariant variables can be found in Sec. {\ref{sec:LNDIS}}. }
\label{fig:invariant-kine}
\end{figure}

Fig. \ref{fig:final-state-kine} shows the distributions of $\theta$ angle, $\phi$ angle,
energy, and pseudorapidity of the final-state electron and neutron.
Note that in the simulation the $z$ direction is defined as the momentum
of the incoming proton beam.
All the scattered electrons go to the central region of the spectrometer ($|\eta|<3$),
and they are precisely and efficiently measured with the central tracker
and electro-magnetic calorimeter \cite{EicCWhitePaperChinese}.
The final neutrons go to very small angles with the pseudorapidity around 5.
They are suggested to be detected with the far-forward neutron calorimeter,
the so called zero-degree calorimeter (ZDC).
Fig. \ref{fig:invariant-kine} shows the cross-section weighted distributions of
the invariant kinematics of interests. We see that most of the events distributed
in the low $Q^2$, small $x_{\rm\pi}$, small $y$ and small $t$ region.
The small-$x_{\pi}$ region is a unique region where EicC can provide the precise data
filling the gap of the current data acquired from the facilities decades ago.
The broad $x_{\rm\pi}$ range from 0.01 to 1 and the high luminosity
of EicC will provide a great opportunity to cross check the large-$x$ behavior
of pion parton distribution when $x_{\rm\pi} \rightarrow 1$,
with the Drell-Yan measurements decades ago.

\section{Statistical error projections of pionic structure function for EicC}
\label{sec:F2piErrors}

\begin{figure}[htbp]
\centering
\includegraphics[scale=0.39]{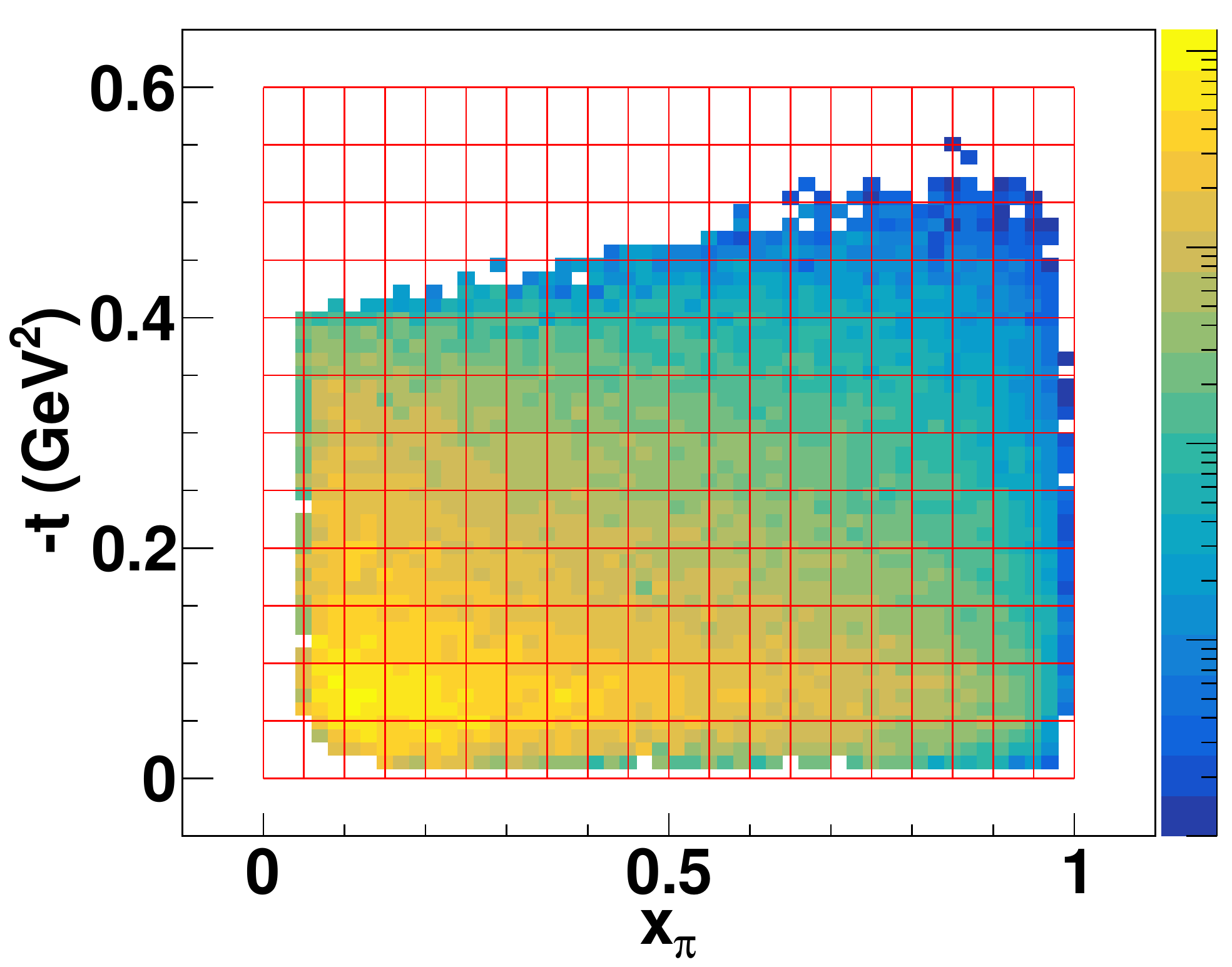}
\caption{The binning scheme in the $x_{\rm\pi}$ vs. $-t$ plane,
for 3 GeV$^2$ $<Q^2<$ 5 GeV$^2$, $x_{\rm L}>0.75$, $P_{\rm T}^{\rm n}<0.5$ GeV,
$M_{\rm X}>0.5$ GeV, and $W^2>4$ GeV$^2$. }
\label{fig:KineBinning}
\end{figure}

\begin{figure}[htbp]
\centering
\includegraphics[scale=0.39]{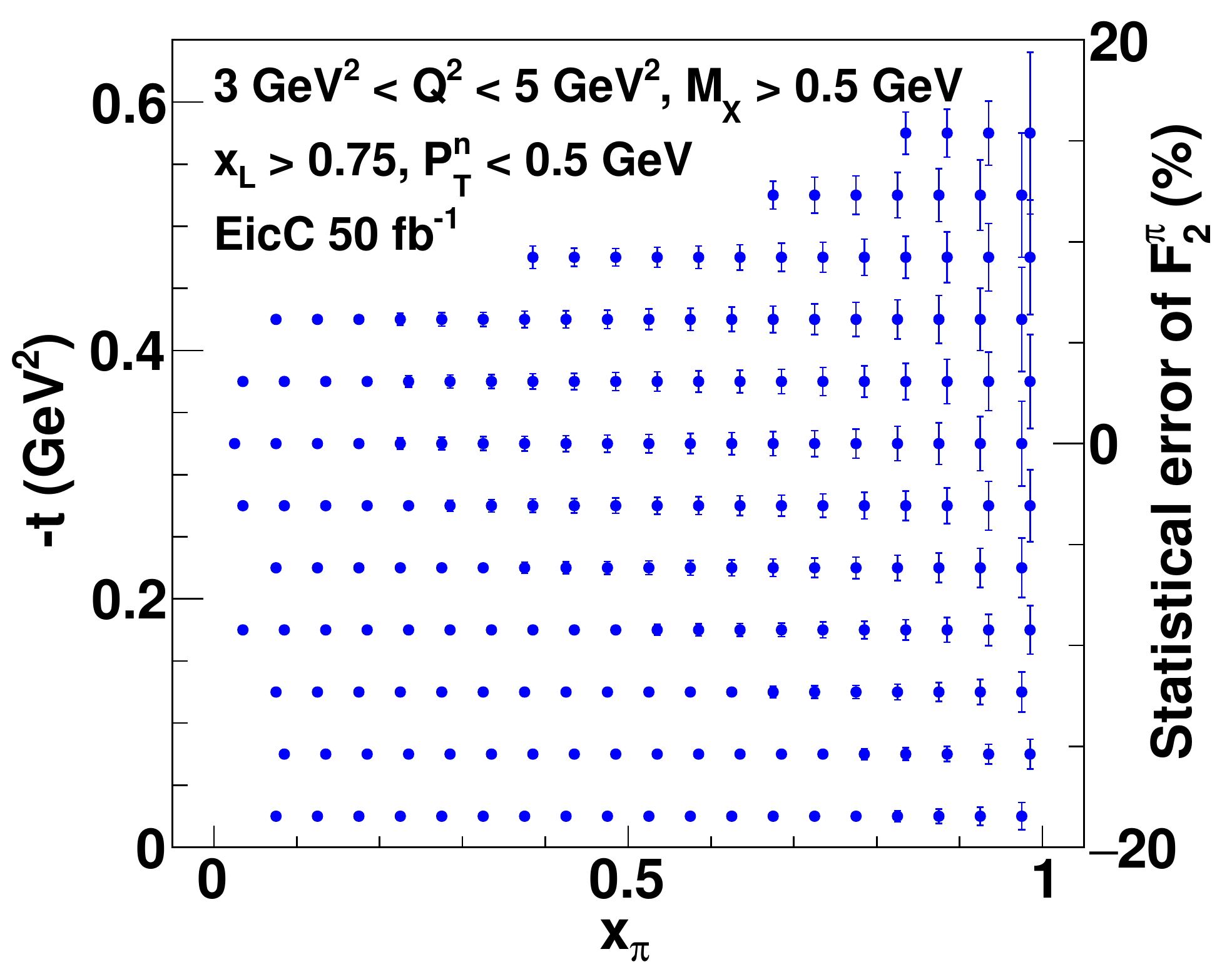}
\caption{The statistical error projections of the
pionic structure function at $Q^2 \sim 4$ GeV$^2$,
for a suggested EicC experiment under an integrated luminosity of 50 fb$^{-1}$.
The left and bottom axes indicate where the bin center of the data point is.
The right axis shows how large the statistical error is. }
\label{fig:F2pi-error-Q2-3-5}
\end{figure}

\begin{figure}[htbp]
\centering
\includegraphics[scale=0.39]{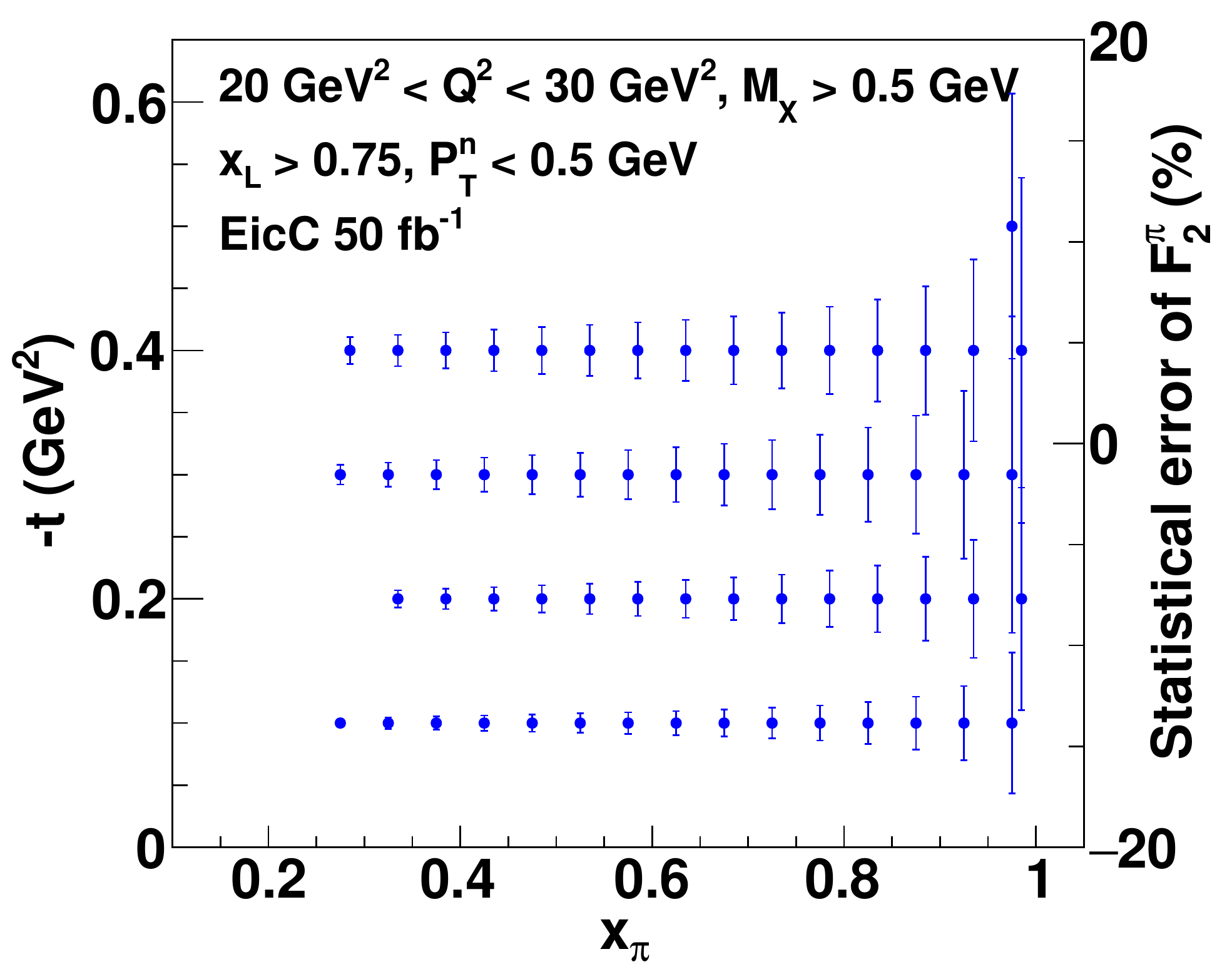}
\caption{The statistical error projections of the
pionic structure function at $Q^2 \sim 25$ GeV$^2$,
for a suggested EicC experiment under an integrated luminosity of 50 fb$^{-1}$.
The left and bottom axes indicate where the bin center of the data point is.
The right axis shows how large the statistical error is. }
\label{fig:F2pi-error-Q2-20-30}
\end{figure}

\begin{figure}[htbp]
\centering
\includegraphics[scale=0.39]{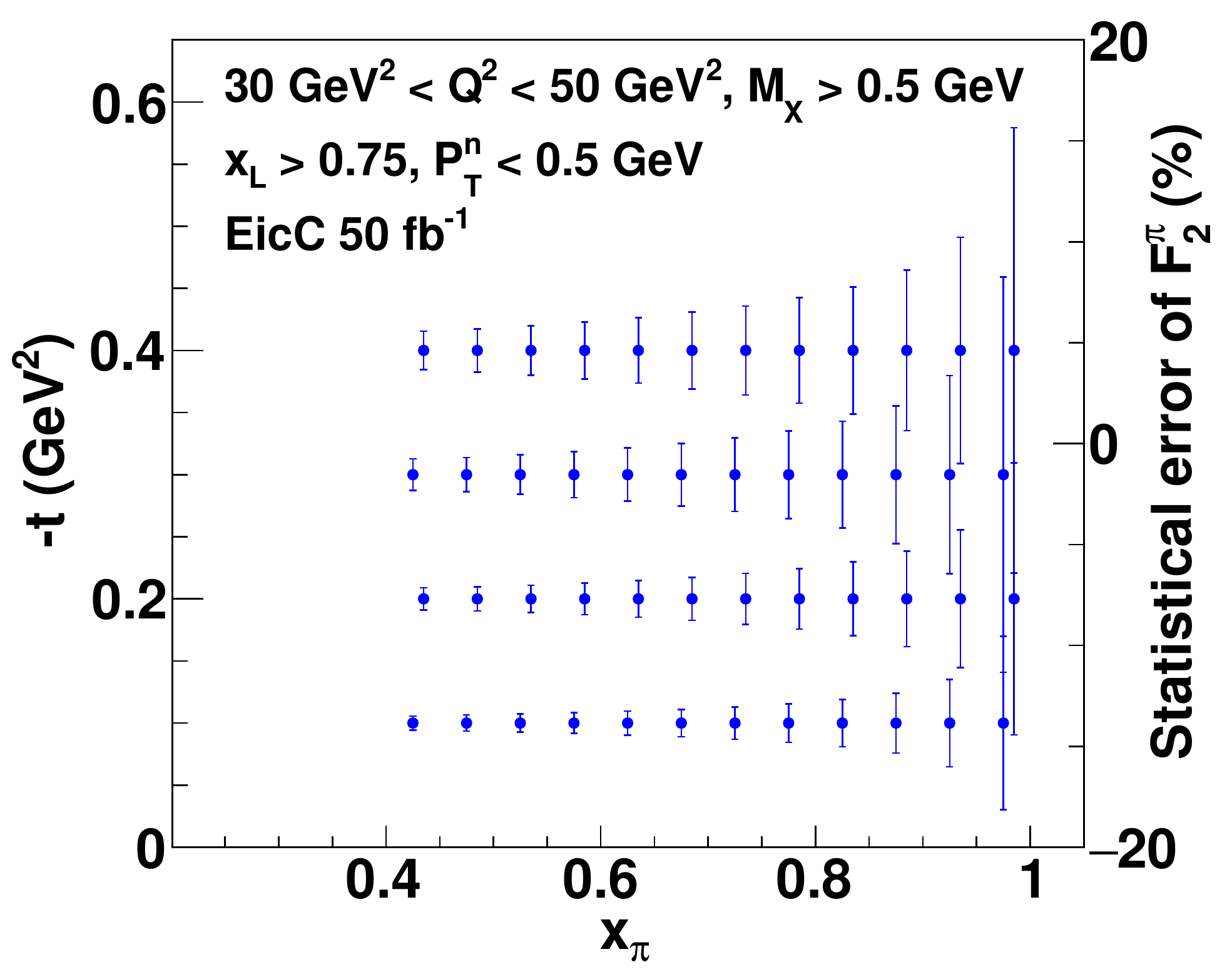}
\caption{The statistical error projections of the
pionic structure function at $Q^2 \sim 40$ GeV$^2$,
for a suggested EicC experiment under an integrated luminosity of 50 fb$^{-1}$.
The left and bottom axes indicate where the bin center of the data point is.
The right axis shows how large the statistical error is. }
\label{fig:F2pi-error-Q2-30-50}
\end{figure}

The statistical uncertainty of the measured experimental observable is directly related
to the number of events collected during an experiment.
Given in the above discussions the calculation of the cross-section,
now we only need to know the accumulated luminosity to estimate the number of events.
we assume the integrated luminosity of an EicC experiment to be 50 fb$^{-1}$,
which corresponds to a run of one to two years.
To study the pionic structure function, we have applied the following
conditions for the event selection: $x_{\rm L}>0.75$, $P_{\rm T}^{\rm n}<0.5$ GeV,
$M_{\rm X}=(p_{\rm e}+P_{\rm p}-p_{\rm e'}-P_{\rm n})^2>0.5$ GeV,
$W>2$ GeV. $x_{\rm L}>0.75$ and $P_{\rm T}^{\rm n}<0.5$ GeV makes sure
the final neutron is a spectator in the e-${\rm \pi}$ DIS process,
where the neutron is from the Fock-state dissociation of the proton,
having a large fraction of the longitudinal momentum of the incoming proton
and a small transverse momentum. $M_{\rm X}>0.5$ GeV requirement is to get rid
of the contamination of e-${\rm \pi}$ elastic scattering process,
and makes sure the struck pion is broke up so as to study the partons inside the pion.
$W>2$ GeV is the common criterion of DIS.

With the above event selection, the LN-DIS events then are divided into different kinematical bins.
Fig. \ref{fig:KineBinning} shows the binning scheme of $x_{\rm \pi}$ and $-t$,
for the low $Q^2$ ($\sim 4$ GeV$^2$) MC data. The number of events in each bin is calculated
with the following formula,
\begin{equation}
N_{\rm i} = \epsilon L\overline{\sigma}_{\rm i} \Delta x_{\rm\pi} \Delta Q^2 \Delta x_{\rm L} \Delta t (1-x_{\rm L}),
\end{equation}
in which $\epsilon$ is the detector efficiency, $L$ is the integrated luminosity,
$\overline{\sigma}_{\rm i}$ is the mean differential cross-section in bin i,
and the rest denotes the sizes of the kinematical bins.
The factor $(1-x_{\rm L})$ is the Jacobian coefficient,
which comes from the transform from integration over $x_{\rm B}$ into integration over $x_{\rm \pi}$.
According to the dimensions and performances of a conceptual design in the far-forward region,
the detector efficiency for neutron can be high. In this simulation,
the efficiency of 50\% is assumed for collecting both the final electron and neutron.
With the number of events in each bin simulated, then the relative statistical error
is estimated to be $1/\sqrt{N_{\rm i}}$.

Fig. \ref{fig:F2pi-error-Q2-3-5} shows the statistical error projections
in a low $Q^2$ bin between 3 GeV$^2$ and 5 GeV$^2$, for an EicC experiment.
The statistical errors are all less than 3\%, starting from $x_{\rm \pi}\sim 0.05$
to $x_{\rm \pi}\sim 1$ at different $t$ bins.
For about half of the data ($x_{\rm\pi}<0.45$), the precisions are very high ($<0.5$\%).
Recalling the uncertainties of pionic parton distributions from the current global analyses
shown in Fig. \ref{fig:Pion_PDF_Sets}, it is very clear that the precision of the EicC measurement
at small $x_{\rm\pi}$ will improve the current analysis tremendously.
The measurement of $t$ is important to know the virtuality of the pion
and to extrapolate the structure function of the real pion.
If we analyze the data at $x_{\rm L}$ around 0.5, we could provide the data of $x_{\rm \pi}$ close to 0.01.
Focusing on the large-$x$ behavior, it is quite exciting to point out that we could measure precisely
the pion structure function of $x_{\rm \pi}$ approaching 0.9.
The error projections of the measurements at high $Q^2$ ($>20$ GeV$^2$) are also projected
and shown in Fig. \ref{fig:F2pi-error-Q2-20-30} and Fig. \ref{fig:F2pi-error-Q2-30-50}.
With fewer bins, the data at high $Q^2$ still possess good precisions.
These precise measurements in different $Q^2$ bins in a broad range of $x_{\rm \pi}$ will
give a test of QCD evolution equations and a better understanding of the gluon distribution of the pion.

\section{Discussions and summary}
\label{sec:summary}

The excellent agreements among the predictions of JAM18 pionic PDFs,
the dynamical parton model predictions \cite{Han:2018wsw,piIMParton-github},
and the H1 data at HERA \cite{Aaron:2010ab} implies
that the LN-DIS process can be used to study the pion structure.
The LN-DIS process can be understood as the scattering between the electron
and the abundant virtual pions around the proton at small momentum transfer $t$.
The pion structure measurement on an electron-ion collider is feasible by tagging the spectator neutron.

Following the pioneering works by H1 and ZEUS, we simulate a LN-DIS experiment at EicC
to investigate the pionic structure function in a wide kinematical range.
The simulation indicates that EicC machine can provide a precise measurement
of the pionic structure of $x_{\rm\pi}$ from 0.01 to 0.9, and of $Q^2$ from 1 GeV$^2$ to 50 GeV$^2$.
Since the neutron is not charged, the very forward neutron
can be separated from the proton beam with a dipole magnet.
Hence measuring the neutrons at very small angles is not difficult,
as long as the space of the tunnel for the accelerator is long enough to install a neutron calorimeter.
In the simulation, we choose a conservative neutron efficiency of 50\%
to model the performance of the neutron detector.
The low energy EicC of high luminosity gives us a good opportunity to
see precisely the one-dimensional structure of the meson.

The pion structure experiment is also proposed at a high energy electron-ion collider in US \cite{Aguilar:2019teb}.
The expected precision of the pion structure data at US-EIC is argued
to be around the same order-of-magnitude in statistics compared to the HERA proton data.
The anticipated error is smaller than 2\% at $x_{\rm\pi}\sim 0.01$ and $Q^2\sim 10$ GeV$^2$
from a simple extraction of the pionic parton distributions  \cite{Aguilar:2019teb}.
The detector capabilities at EicC and US-EIC will be very similar.
The US-EIC will be running at a much higher energy and
the high statistical data will be mainly in the small-$x_{\rm \pi}$ region ($\lesssim 0.3$).
In the large-$x_{\rm \pi}$ region ([0.3, 0.9]), EicC has the competing quality of the pion structure data.
Therefore EicC will play an important role in the synergy of exploring the meson structure.

The precise measurement of the pionic structure function in a broad kinematical domain
definitely will flourish our understanding of the strong interaction.
The systematical study of the LN-DIS reaction and the precise extraction
of the pionic structure from sea quark region to valence quark region at EicC will
be a critical input for the database of PDFs of hadrons.
The PDFs on the experimental side will differentiate
the various models on hadron structure we have so far.
Most importantly, the LN-DIS experiment at EicC has a great potential
to reveal pion parton distributions with a lot of details,
leading to a better understanding of many nonperturbative approaches,
the dynamical symmetry breaking, and why the pion mass is so small
compared to the proton mass.

\begin{acknowledgments}
Sincere acknowledgment is given to Dr. Jixie ZHANG for so many friendly helps
and the cross-checks of the simulation results with an independent estimation by him.
We are very grateful for Craig ROBERTS' suggestions on studying the pionic structure at EicC
and his efforts on pushing the project. R. WANG expresses his gratitude to Tanja HORN
for her professional and constructive questions on the experimental side.
We thank the EicC working group for the suggestions and the fruitful discussions.
This work is supported by the Strategic Priority Research Program of Chinese Academy of Sciences
under the Grant NO. XDB34030301.
\end{acknowledgments}

\bibliographystyle{apsrev4-1}
\bibliography{refs}

\end{document}